\documentclass[reprint,amsmath,amssymb,aps,floatfix,superscriptaddress,showpacs,prl]{revtex4-1}
\usepackage{graphicx}
\usepackage{dcolumn}
\usepackage{bm}
\usepackage{color}
\usepackage{subfigure}
\usepackage{url}

\usepackage{comment}

\newcommand{\INFNFI}{INFN Sezione di Florence, Via Sansone, 1 - 50019, Sesto Fiorentino, Italy}
\newcommand{\UNIFI}{Department of Physics, University of Florence, Via Sansone, 1 - 50019, Sesto Fiorentino, Italy}
\newcommand{\WasedaUniv}{Waseda Research Institute for Science and Engineering, Waseda University, 17 Kikuicho,  Shinjuku, Tokyo 162-0044, Japan}
\newcommand{\JEM}{JEM Utilization Center, Human Spaceflight Technology Directorate, Japan Aerospace Exploration Agency, 2-1-1 Sengen, Tsukuba, Ibaraki 305-8505, Japan}
\newcommand{\ICRR}{Institute for Cosmic Ray Research, The University of Tokyo, 5-1-5 Kashiwa-no-Ha, Kashiwa, Chiba 277-8582, Japan}
\newcommand{\UNISI}{Department of Physical Sciences, Earth and Environment, University of Siena, via Roma 56, 53100 Siena, Italy}
\newcommand{\INFNPI}{INFN Sezione di Pisa, Polo Fibonacci, Largo B. Pontecorvo, 3 - 56127 Pisa, Italy}
\newcommand{\WashingtonUniv}{Department of Physics and McDonnell Center for the Space Sciences, Washington University, One Brookings Drive, St. Louis, Missouri 63130-4899, USA}
\newcommand{\HelPhysLab}{Heliospheric Physics Laboratory, NASA/GSFC, Greenbelt, Maryland 20771, USA}
\newcommand{\MarylandUniv}{Center for Space Sciences and Technology, University of Maryland, Baltimore County, 1000 Hilltop Circle, Baltimore, Maryland 21250, USA}
\newcommand{\CRESSTMaryland}{Center for Research and Exploration in Space Sciences and Technology, NASA/GSFC, Greenbelt, Maryland 20771, USA}
\newcommand{\AstrPhysLab}{Astroparticle Physics Laboratory, NASA/GSFC, Greenbelt, Maryland 20771, USA}
\newcommand{\IFAC}{Institute of Applied Physics (IFAC),  National Research Council (CNR), Via Madonna del Piano, 10, 50019, Sesto Fiorentino, Italy}
\newcommand{\LousianaUniv}{Department of Physics and Astronomy, Louisiana State University, 202 Nicholson Hall, Baton Rouge, Louisiana 70803, USA}
\newcommand{\UNIPD}{Department of Physics and Astronomy, University of Padova, Via Marzolo, 8, 35131 Padova, Italy}
\newcommand{\INFNPD}{INFN Sezione di Padova, Via Marzolo, 8, 35131 Padova, Italy}
\newcommand{\JAXA}{Institute of Space and Astronautical Science, Japan Aerospace Exploration Agency, 3-1-1 Yoshinodai, Chuo, Sagamihara, Kanagawa 252-5210, Japan}
\newcommand{\KanagawaUniv}{Kanagawa University, 3-27-1 Rokkakubashi, Kanagawa, Yokohama, Kanagawa 221-8686, Japan}
\newcommand{\HirosakiUniv}{Faculty of Science and Technology, Graduate School of Science and Technology, Hirosaki University, 3, Bunkyo, Hirosaki, Aomori 036-8561, Japan}
\newcommand{\KyotoUniv}{Yukawa Institute for Theoretical Physics, Kyoto University, Kitashirakawa Oiwake-cho, Sakyo-ku, Kyoto, 606-8502, Japan}
\newcommand{\DepElecInfSys}{Department of Electronic Information Systems, Shibaura Institute of Technology, 307 Fukasaku, Minuma, Saitama 337-8570, Japan}
\newcommand{\WasedaShool}{School of Advanced Science and Engineering, Waseda University, 3-4-1 Okubo, Shinjuku, Tokyo 169-8555, Japan}
\newcommand{\NatInstPolRes}{National Institute of Polar Research, 10-3, Midori-cho, Tachikawa, Tokyo 190-8518, Japan}
\newcommand{\YokohamaUniv}{Faculty of Engineering, Division of Intelligent Systems Engineering, Yokohama National University, 79-5 Tokiwadai, Hodogaya, Yokohama 240-8501, Japan}
\newcommand{\ShinshuUniv}{Faculty of Science, Shinshu University, 3-1-1 Asahi, Matsumoto, Nagano 390-8621, Japan}
\newcommand{\DepAstrKyotoUniv}{Department of Astronomy, Graduate School of Science, Kyoto University, Kitashirakawa Oiwake-cho, Sakyo-ku, Kyoto, 606-8502, Japan}
\newcommand{\InstPartNuclStudTsukuba}{Institute of Particle and Nuclear Studies, High Energy Accelerator Research Organization, 1-1 Oho, Tsukuba, Ibaraki, 305-0801, Japan}
\newcommand{\UNIPI}{University of Pisa, Polo Fibonacci, Largo B. Pontecorvo, 3 - 56127 Pisa, Italy}
\newcommand{\IbarakiCollege}{Department of Electrical and Electronic Systems Engineering, National Institute of Technology, Ibaraki College, 866 Nakane, Hitachinaka, Ibaraki 312-8508, Japan}
\newcommand{\DepAstrMarylandUniv}{Department of Astronomy, University of Maryland, College Park, Maryland 20742, USA}
\newcommand{\RitsumeikanUniv}{Department of Physical Sciences, College of Science and Engineering, Ritsumeikan University, Shiga 525-8577, Japan}
\newcommand{\WasedaFaculty}{Faculty of Science and Engineering, Global Center for Science and Engineering, Waseda University, 3-4-1 Okubo, Shinjuku, Tokyo 169-8555, Japan}
\newcommand{\DenverUniv}{Department of Physics and Astronomy, University of Denver, Physics Building, Room 211, 2112 East Wesley Avenue, Denver, Colorado 80208-6900, USA}
\newcommand{\QuantumICT}{Quantum ICT Advanced Development Center, National Institute of Information and Communications Technology, 4-2-1 Nukui-Kitamachi, Koganei, Tokyo 184-8795, Japan}
\newcommand{\AoyamaUniv}{College of Science and Engineering, Department of Physics and Mathematics, Aoyama Gakuin University,  5-10-1 Fuchinobe, Chuo, Sagamihara, Kanagawa 252-5258, Japan}
\newcommand{\NihonUniv}{College of Industrial Technology, Nihon University, 1-2-1 Izumi, Narashino, Chiba 275-8575, Japan}
\newcommand{\OsakaUnivDMP}{Division of Mathematics and Physics, Graduate School of Science, Osaka City University, 3-3-138 Sugimoto, Sumiyoshi, Osaka 558-8585, Japan}
\newcommand{\OsakaUnivNYITEP}{Nambu Yoichiro Institute of Theoretical and Experimental Physics, Osaka City University, 3-3-138 Sugimoto, Sumiyoshi, Osaka 558-8585, Japan}
\newcommand{\NIQRSTAnagawa}{National Institutes for Quantum and Radiation Science and Technology, 4-9-1 Anagawa, Inage, Chiba 263-8555, Japan}
\newcommand{\NagoyaUniv}{Nagoya University, Furo, Chikusa, Nagoya 464-8601, Japan}
\newcommand{\CSIbarakiUniv}{College of Science, Ibaraki University, 2-1-1 Bunkyo, Mito, Ibaraki 310-8512, Japan}

\usepackage[colorlinks]{hyperref}
\hypersetup{bookmarks, colorlinks=true, citecolor=cyan, filecolor=blue ,linkcolor=red, urlcolor=blue}

\begin{document}


\title{Direct Measurement of the Nickel Spectrum in Cosmic Rays in the Energy Range\\
	from 8.8 GeV$/n$ to 240 GeV$/n$ 
  with CALET on the International Space Station}

\author{O.~Adriani}
\affiliation{\UNIFI}
\affiliation{\INFNFI}
\author{Y.~Akaike}
\email[Corresponding author: ]{yakaike@aoni.waseda.jp}
\affiliation{\WasedaUniv}
\affiliation{\JEM}
\author{K.~Asano}
\affiliation{\ICRR}
\author{Y.~Asaoka}
\affiliation{\ICRR}
\author{E.~Berti} 
\affiliation{\UNIFI}
\affiliation{\INFNFI}
\author{G.~Bigongiari}
\email[Corresponding author: ]{bigongiari2@unisi.it}
\affiliation{\UNISI}
\affiliation{\INFNPI}
\author{W.~R.~Binns}
\affiliation{\WashingtonUniv}
\author{M.~Bongi}
\affiliation{\UNIFI}
\affiliation{\INFNFI}
\author{P.~Brogi}
\affiliation{\UNISI}
\affiliation{\INFNPI}
\author{A.~Bruno}
\affiliation{\HelPhysLab}
\author{J.~H.~Buckley}
\affiliation{\WashingtonUniv}
\author{N.~Cannady}
\affiliation{\MarylandUniv}
\affiliation{\AstrPhysLab}
\affiliation{\CRESSTMaryland}
\author{G.~Castellini}
\affiliation{\IFAC}
\author{C.~Checchia}
\email[Corresponding author: ]{caterina.checchia2@unisi.it}
\affiliation{\UNISI}
\affiliation{\INFNPI}
\author{M.~L.~Cherry}
\affiliation{\LousianaUniv}
\author{G.~Collazuol}
\affiliation{\UNIPD}
\affiliation{\INFNPD} 
\author{K.~Ebisawa}
\affiliation{\JAXA}
\author{A.~W.~Ficklin}
\affiliation{\LousianaUniv}
\author{H.~Fuke}
\affiliation{\JAXA}
\author{S.~Gonzi}
\affiliation{\UNIFI}
\affiliation{\INFNFI}
\author{T.~G.~Guzik}
\affiliation{\LousianaUniv}
\author{T.~Hams}
\affiliation{\MarylandUniv}
\author{K.~Hibino}
\affiliation{\KanagawaUniv}
\author{M.~Ichimura}
\affiliation{\HirosakiUniv}
\author{K.~Ioka}
\affiliation{\KyotoUniv}
\author{W.~Ishizaki}
\affiliation{\ICRR}
\author{M.~H.~Israel}
\affiliation{\WashingtonUniv}
\author{K.~Kasahara}
\affiliation{\DepElecInfSys}
\author{J.~Kataoka}
\affiliation{\WasedaShool}
\author{R.~Kataoka}
\affiliation{\NatInstPolRes}
\author{Y.~Katayose}
\affiliation{\YokohamaUniv}
\author{C.~Kato}
\affiliation{\ShinshuUniv}
\author{N.~Kawanaka}
\affiliation{\KyotoUniv}
\affiliation{\DepAstrKyotoUniv}
\author{Y.~Kawakubo}
\affiliation{\LousianaUniv}
\author{K.~Kobayashi}
\affiliation{\WasedaUniv}
\affiliation{\JEM}
\author{K.~Kohri} 
\affiliation{\InstPartNuclStudTsukuba} 
\author{H.~S.~Krawczynski}
\affiliation{\WashingtonUniv}
\author{J.~F.~Krizmanic}
\affiliation{\AstrPhysLab}
\author{P.~Maestro}
\affiliation{\UNISI}
\affiliation{\INFNPI}
\author{P.~S.~Marrocchesi}
\affiliation{\UNISI}
\affiliation{\INFNPI}
\author{A.~M.~Messineo}
\affiliation{\UNIPI}
\affiliation{\INFNPI}
\author{J.~W.~Mitchell}
\affiliation{\AstrPhysLab}
\author{S.~Miyake}
\affiliation{\IbarakiCollege}
\author{A.~A.~Moiseev}
\affiliation{\DepAstrMarylandUniv}
\affiliation{\AstrPhysLab}
\affiliation{\CRESSTMaryland}
\author{M.~Mori}
\affiliation{\RitsumeikanUniv}
\author{N.~Mori}
\affiliation{\INFNFI}
\author{H.~M.~Motz}
\affiliation{\WasedaFaculty}
\author{K.~Munakata}
\affiliation{\ShinshuUniv}
\author{S.~Nakahira}
\affiliation{\JAXA}
\author{J.~Nishimura}
\affiliation{\JAXA}
\author{G.~A.~de~Nolfo}
\affiliation{\HelPhysLab}
\author{S.~Okuno}
\affiliation{\KanagawaUniv}
\author{J.~F.~Ormes}
\affiliation{\DenverUniv}
\author{N.~Ospina}
\affiliation{\UNIPD}
\affiliation{\INFNPD} 
\author{S.~Ozawa}
\affiliation{\QuantumICT}
\author{L.~Pacini}
\affiliation{\UNIFI}
\affiliation{\IFAC}
\affiliation{\INFNFI}
\author{P.~Papini}
\affiliation{\INFNFI}
\author{B.~F.~Rauch}
\affiliation{\WashingtonUniv}
\author{S.~B.~Ricciarini}
\affiliation{\IFAC}
\affiliation{\INFNFI}
\author{K.~Sakai}
\affiliation{\MarylandUniv}
\affiliation{\AstrPhysLab}
\affiliation{\CRESSTMaryland}
\author{T.~Sakamoto}
\affiliation{\AoyamaUniv}
\author{M.~Sasaki}
\affiliation{\DepAstrMarylandUniv}
\affiliation{\AstrPhysLab}
\affiliation{\CRESSTMaryland}
\author{Y.~Shimizu}
\affiliation{\KanagawaUniv}
\author{A.~Shiomi}
\affiliation{\NihonUniv}
\author{P.~Spillantini}
\affiliation{\UNIFI}
\author{F.~Stolzi}
\email[Corresponding author: ]{francesco.stolzi@unisi.it}
\affiliation{\UNISI}
\affiliation{\INFNPI}
\author{S.~Sugita}
\affiliation{\AoyamaUniv}
\author{A.~Sulaj} 
\affiliation{\UNISI}
\affiliation{\INFNPI}
\author{M.~Takita}
\affiliation{\ICRR}
\author{T.~Tamura}
\affiliation{\KanagawaUniv}
\author{T.~Terasawa}
\affiliation{\ICRR}
\author{S.~Torii}
\affiliation{\WasedaUniv}
\author{Y.~Tsunesada}
\affiliation{\OsakaUnivDMP}
\affiliation{\OsakaUnivNYITEP}
\author{Y.~Uchihori}
\affiliation{\NIQRSTAnagawa}
\author{E.~Vannuccini}
\affiliation{\INFNFI}
\author{J.~P.~Wefel}
\affiliation{\LousianaUniv}
\author{K.~Yamaoka}
\affiliation{\NagoyaUniv}
\author{S.~Yanagita}
\affiliation{\CSIbarakiUniv}
\author{A.~Yoshida}
\affiliation{\AoyamaUniv}
\author{K.~Yoshida}
\affiliation{\DepElecInfSys}
\author{W.~V.~Zober}
\affiliation{\WashingtonUniv}

\collaboration{CALET Collaboration}

\date{\today}

\begin{abstract}
	The relative abundance of cosmic ray nickel nuclei with respect to iron is by far larger than for all other trans-iron elements, therefore it provides a favorable opportunity for a low background measurement of its spectrum. Since nickel, as well as iron, is one of the most stable nuclei, the nickel energy spectrum and its relative abundance with respect to iron provide important information to estimate the abundances at the cosmic ray source and to model the Galactic propagation of heavy nuclei.
However, only a few direct measurements of cosmic-ray nickel at energy larger than $ \sim 3~\mathrm{GeV}/n $ are available at present in the literature and they are affected by strong limitations in both energy reach and statistics.  In this paper we present a measurement of the differential energy spectrum of nickel in the energy range from 8.8 to 240~GeV$ /n $, carried out with unprecedented precision by the Calorimetric Electron Telescope (CALET) in operation on the International Space Station since 2015. The CALET instrument can identify individual nuclear species via a measurement of their electric charge with a dynamic range extending far beyond iron (up to atomic number $ Z $ = 40). The particle's energy is measured by a homogeneous calorimeter (1.2 proton interaction lengths, 27~radiation lengths) preceded by a thin  imaging section (3 radiation lengths) providing tracking and energy sampling.
This paper follows our previous measurement of the iron spectrum~[O.~Adriani $ et$~$ al. $ (CALET Collaboration), Phys. Rev. Lett. 126, 241101 (2021).], and it extends our investigation on the energy dependence of the spectral index of heavy elements. It reports the analysis of nickel data collected from November 2015 to May 2021 and a detailed assessment of the systematic uncertainties. 
In the region from 20 to 240~GeV$ /n $ our present data are compatible within the errors with a single power law with spectral index $ -2.51 \pm 0.07 $.  

\end{abstract}

\maketitle

\section{Introduction}
The energy spectra and relative abundances of cosmic rays (CR) are key observables for a theoretical understanding of the acceleration and propagation mechanisms of charged particles in our Galaxy~\cite{Drury, Ohira2011,Malkov,Blasi2012,Tomassetti,Vladimirov,Ptuskin,Thoudam,Bernard,Serpico,Ohira2016, Evoli2018,Evoli2019,ICRC2021_Caprioli,ICRC2021_Lipari}.  
Direct measurements by space-borne instruments have recently achieved a level of unprecedented precision, thanks to long term observations and their capability to identify individual chemical elements. Direct measurements from high-altitude balloons and indirect measurements from ground based arrays convey important complementary information, albeit with different systematic uncertainties.
 The extensions to higher energies of CR spectral data have shown unexpected deviations from a single power law, as in the case of the recently observed double broken spectral shape of the proton spectrum in the multi-TeV domain, reported by the DAMPE~\cite{DAMPE_proton} and Calorimetric Electron Telescope (CALET)~\cite{ICRC2021_KKPSM} experiment. A progressive spectral hardening (as a function of energy) has been established for light elements and heavier nuclei~\cite{AMS-CO, AMS-Li-Be-B, AMS-Ne-Mg-Si, CREAM2HARD, CALET-CO} with an onset at a few hundred GeV$/n$. Also,  a spectral softening has been observed in the TeV domain for proton and helium as reported by the DAMPE~\cite{DAMPE_proton,DAMPE-He}, CALET~\cite{ICRC2021_KKPSM} and  NUCLEON~\cite{NUCLEON-nuclei} experiments.
The spectral study of
heavy elements was recently extended to higher energies with the publication of the iron spectrum by the AMS-02~\cite{AMS-Fe} and CALET~\cite{CALET-IRON2021} experiments. \\
\indent
 In this paper, we pursue the study of elements sitting on the high side of the periodic table, where nickel -- with much larger abundance
 than all other trans-iron elements -- provides a favorable opportunity for a low background measurement of its spectrum. 
 \\
\indent
Space-borne direct measurements of CR nickel nuclei include the spectrum measured from $0.6$ to 35~GeV/${n}$ by~the French-Danish C2 instrument HEAO3-C2~\cite{HEAO3-Ni} onboard the NASA HEAO3 satellite (launched in 1979)~and the recent measurement by the NUCLEON experiment (launched in 2014)~\cite{NUCLEON-Ni} in the energy range~$\sim$51--511~GeV$/n$.
Measurements in the lower energy range~50--550~MeV/${n}$ were carried out, during the 2009--2010 solar minimum period, by the Cosmic Ray Isotope Spectrometer (CRIS)~\cite{Lave13} onboard the Advanced Composition Explorer at the L1 Lagrange point. 
Data up to $\sim$~500~MeV/${n}$~\cite{Cummings2016} were collected by the $\it{Voyager\, 1}$ spacecraft after the start of observations of the local interstellar energy spectra of Galactic cosmic-ray nuclei (August 2012).\\
\indent
Earlier measurements with balloon experiments have a limited statistics and energy reach. They include
(i) the High Energy Nuclei (HEN) telescope~\cite{Juliusson74} at nickel energies up to about 10.5~GeV/${n}$ (3 flights in 1971 and 1972 for a total of 7.6 m$^{2}$ sr hrs);
(ii) the scintillation-Cherenkov telescope (hereafter cited as Balloon 1975)~\cite{Minagawa81} from 1 to 10~GeV$ /n $ (2 flights in 1975 with a total exposure of 20 m$^{2}$ sr hrs); 
(iii) the multi-element Cherenkov telescope~\cite{Lezniak78}
  from 0.3--50~GeV$/n$ (3 flights in 1974 and one in 1976);  
(iv) the Cosmic Ray Isotope Instrument System (CRISIS)~\cite{Young81} from 600--900~MeV/${n}$ ($\sim$~57 hrs afloat in 1975);
(v) the Large Isotopic Composition Experiment (ALICE)~\cite{Esposito92} at energies near 1~GeV/${n}$ (flown for 14.7 hrs in 1987).
\\
\indent
In this paper we present
a measurement of the differential energy spectrum of CR nickel in the energy range from 8.8 to 240~GeV$ /n $ carried out, with unprecedented precision, with CALET onboard the International Space Station (ISS).
Though optimized for the measurement
of the all-electron spectrum~\cite{CALET-ELE2017,CALET-ELE2018},
CALET has an excellent charge identification capability to tag individual CR elements~\cite{CALET-PROTON,CALET, CALET2, CALET3} from proton to nickel nuclei (and above). It can explore particle energies up to the PeV scale thanks to its large dynamic range, adequate calorimetric depth and accurate tracking.
CALET published accurate spectral measurements of electrons \cite{CALET-ELE2018}, protons~\cite{CALET-PROTON}, carbon~\cite{CALET-CO}, oxygen~\cite{CALET-CO}, and iron~\cite{CALET-IRON2021}. Preliminary updates of proton, helium, boron and boron to carbon ratio analyses were presented at the ICRC-2021 conference~\cite{ICRC2021_HL}. 
\looseness=1

\section{CALET Instrument}
Charge identification is carried out by the CHarge Detector (CHD), a two-layered hodoscope of plastic scintillator paddles.
It can resolve individual elements from atomic number $ Z $~=~1 to $ Z $~=~40 with excellent charge resolution spanning from 0.15 charge units for C to 0.35 charge units for Fe~\cite{GSI}. The particle's energy is measured with the Total AbSorption Calorimeter (TASC), a lead-tungstate homogeneous calorimeter [27 radiation lengths (r.l.), 1.2 proton interaction lengths] preceded by a thin (3 r.l.) pre-shower IMaging Calorimeter (IMC).
The latter is equipped with 16 layers of thin scintillating fibers (1 $ \mathrm{mm}^2$ square cross-section) read out individually and interleaved with tungsten absorbers. The IMC provides tracking capabilities as well as an independent charge measurement, via multiple samples of specific energy loss ($ dE/dx $) in each fiber, up to the onset of saturation which occurs for ions above silicon. Therefore charge identification for nickel and neighboring elements relies on CHD only.
More details on the instrument and on the trigger system can be found in the Supplemental Material (SM) of Ref.~\cite{CALET-ELE2017}.
CALET was launched on August 19, 2015 and installed on the Japanese Experiment Module Exposed Facility of the ISS. The on-orbit commissioning phase was successfully completed in the first days of October 2015.
Calibration and test of the instrument took place at the CERN-SPS during five campaigns between 2010 and 2015 with beams of electrons, protons and relativistic ions~\cite{akaike2015, bigo, niita}.

\section{Data analysis}
The flight data (FD) used in the present analysis were collected over a period of 2038 
 days of CALET operation. 
The total observation live time for  the high-energy (HE) shower trigger~\cite{CALET2017} is $T\sim 4.1 \times10^4$ hours, corresponding to 86.0\% of total observation time. 

Individual on-orbit calibration of all channels is performed with a dedicated trigger mode~\cite{CALET2017,niita} allowing the selection of penetrating protons and He particles.  First, raw data are corrected for gain differences among the channels,  light output non-uniformity and any residual dependence on time and temperature. After calibration, a single ``best track'' is reconstructed for each event with an associated estimate of its charge and energy.

The particle's direction and entrance point are reconstructed from the coordinates of the scintillating fibers in the IMC. The tracking algorithm, based on a combinatorial Kalman filter, identifies the incident track in the presence of background hits generated by backscattered radiation from the TASC~\cite{paolo2017}.
The angular resolution  and the spatial resolution for the impact point on the CHD  are $\sim{0.08}^\circ$  and $\sim$180 $\mu$m respectively.

Physics processes and interactions in the apparatus are simulated via Monte Carlo (MC) techniques, based on the EPICS package~\cite{EPICS, EPICSurl} which implements the hadronic interaction model DPMJET-III~\cite{dpmjet3prl}. The instrument configuration and detector response are detailed in the simulation code which provides digitized signals from all channels. An independent analysis based on GEANT4~\cite{GEANT4} is also performed to assess the systematic uncertainties. In this analysis, only the $ ^{58}\mathrm{Ni} $ isotope was considered since its mass difference with respect to other isotopes (mainly $ ^{60}\mathrm{Ni} $) is less than 3\%.

\indent
\emph{\bf{Charge measurement}}
\indent
The particle's charge $ Z $ is reconstructed 
from the signals of the CHD paddles traversed by the incident particle and properly corrected for its path length.
Either CHD layer provides an independent $ dE/dx $ measurement which has to be corrected for the
 quenching effect in the scintillator's light yield. The latter is parameterized by fitting selected FD samples of each nuclear species to a ``halo'' model~\cite{GSI} as a function of $ Z^2 $.  
The resulting curves are then used to reconstruct a charge value in either layer ($Z_{\rm CHDX}$, $Z_{\rm CHDY}$) on an event-by-event basis~\cite{CALET-CO}.
The presence of an increasing amount of backscatters from the TASC at higher energy generates additional energy deposits in the CHD that add on to the primary particle ionization signal and may induce a wrong charge identification.
This effect causes a systematic displacement of the CHDX/CHDY charge peaks to higher values (up to 0.8 charge units) with respect to the nominal charge position. Therefore it is necessary to restore the nickel peak position to its nominal value, $ Z $~=~28,  by an energy dependent charge correction applied separately to the FD and the MC data. A similar correction is applied to iron and nearby elements.
The CHD charge resolution $ \sigma_Z $, obtained by combining the average of $Z_{\rm CHDX}$ and $Z_{\rm CHDY}$ signal is 0.39 in charge units and it is shown in Fig.~S1 of the SM~\cite{PRL-SM}. Background contamination from neighbor elements misidentified as nickel is shown in Fig.~S2 of the SM~\cite{PRL-SM}. Between 100~GeV and 1~TeV it is mainly due to iron and secondly to cobalt. Above 1~TeV the iron contribution is the most important. Contamination from heavier nuclei is negligible.

\begin{figure} [!htb]
\centering
\includegraphics[width=1.0\hsize]{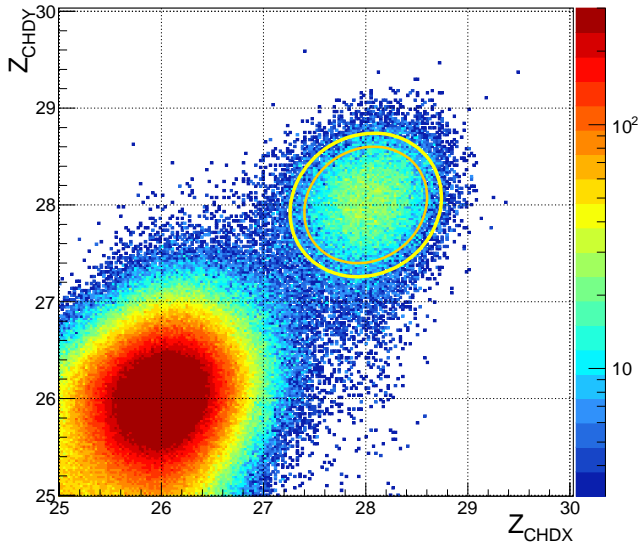}
\caption{\scriptsize Crossplot of $Z_{\rm CHDY}$ vs.~$Z_{\rm CHDX}$  reconstructed charges in the elemental range between Mn ($ Z $~=~25) and Zn ($ Z $~=~30) before removing charge-changing nuclear interactions. Nickel candidates are selected inside an ellipse with semi-minor and major axes 1.4 $ \sigma_x $ and 1.4 $  \sigma_y $, respectively, rotated clockwise by $ 45^{\circ} $. The maximum and the minimum elliptical selection (depending on the energy) are indicated by the yellow and the orange ellipses in the figure.} 
\label{fig:CrossPlotCharge}
\end{figure}\noindent

\indent
\emph{\bf{Energy measurement}}
\indent
For each event, the shower energy $E_{\rm TASC}$  is calculated as
the sum of the energy deposits of all TASC logs, after merging the calibrated gain ranges of each channel~\cite{CALET2017}.
The energy response derived from the MC simulations was tuned
using the results of a beam test carried out at the CERN-SPS in 2015~\cite{akaike2015}
with beams of accelerated ion fragments  of 13, 19 and 150~GeV$ /c/n $ momentum per nucleon (as described in the SM of Ref.~\cite{CALET-CO}).
Correction factors are 6.7\% for  $E_{\rm TASC}<45$~GeV and 3.5\% for $E_{\rm TASC}>350$~GeV, respectively. A linear interpolation is used to determine the correction factor for intermediate energies.

\indent
\emph{\bf{Event selection}}
\indent
The onboard HE shower trigger,  
based on the coincidence of the summed dynode signals of the last four IMC layers and the top TASC layer (TASCX1) is fully efficient for elements heavier than oxygen.
Therefore, an offline trigger confirmation, as required for the analysis of lower charge elements~\cite{CALET-PROTON,CALET-CO}, is not necessary for nickel, because the HE trigger threshold is far below the signal amplitude expected from a nickel ion at minimum ionization (MI) and the trigger efficiency is close to 100\%. However, in order to select interacting particles, a deposit larger than 2 standard deviation of the MI peak is required in at least one of the first four layers of the TASC.

Events with one well-fitted track crossing the whole detector from the top of the CHD to the TASC bottom layer 
(and clear from the edges of TASCX1 by at least 2~cm) are selected.
The fiducial geometrical factor for this category of events is $S\Omega$~$\sim$~$510 \, \mathrm{cm}^2$sr, corresponding to about 50\% of the CALET total acceptance.

Particles undergoing a charge-changing nuclear interaction in the upper part of the instrument 
are removed by requiring that the difference between the charges from either layer of the CHD is less than $1.5 $ charge units. 
 The cross plot of the  $ Z_{\mathrm{CHDY}} $ vs.~$ Z_{\mathrm{CHDX}}$ charge, in Fig.~\ref{fig:CrossPlotCharge}, shows the nickel events selection: candidates are contained within an ellipse centered at $ Z $ = 28 with 1.4~$ \sigma_x $ and 1.4~$ \sigma_y $ wide semi-major and minor axes (with both variances depending on the energy) for $ Z_{\mathrm{CHDX}} $ and $ Z_{\mathrm{CHDY}}$, respectively, and rotated clockwise by 45$ ^{\circ} $. 
Event selections are identical for the MC and the FD.

\indent
\emph{\bf{Energy unfolding}}
\indent
As detailed in Ref.~\cite{CALET-IRON2021} for iron,
the TASC crystals are subject to a light quenching phenomenon which is not reproduced by the MC simulations. Therefore a quenching correction is extracted from the FD and applied \textit{a posteriori} 
to the MC energy deposits generated by non-interacting primary particles in the TASC logs. 
Distributions of $E_{\rm TASC}$ for Ni selected candidates are shown in Fig.~S2 of the SM~\cite{PRL-SM}, with a sample of $5.2  \times 10^3$ events.

In order to take into account the li\-mi\-ted calorimetric energy resolution for hadrons (of the order of $\sim$30\%) an energy unfolding algorithm is applied to correct for bin-to-bin migration effects. 
In this analysis, we used the Bayesian approach~\cite{Ago} 
implemented within the RooUnfold package~\cite{ROOUNFOLD} of the ROOT analysis framework~\cite{ROOT}.
Each element of the response matrix represents
the probability that a primary nucleus in a given energy interval of the CR spectrum produces 
an energy deposit falling into a given bin of $E_{\rm TASC}$.
The response matrix
is derived using the MC simulation after applying the same selection procedure as for flight data and it is shown in Fig.~S6 of the SM~\cite{PRL-SM}.  

\indent
\emph{\bf{Differential energy spectrum}}
\indent
The energy spectrum is obtained from the unfolded energy distribution as follows:
\begin{equation}
\Phi(E) = \frac{N(E)}{\Delta E\;  \varepsilon(E) \;  S\Omega \;  T }
\label{eq_flux}
\end{equation}
\begin{equation}
N(E) = U \left[N_{obs}(E_{\rm TASC}) - N_{bg}(E_{\rm TASC}) \right]
\end{equation}
where $ S\Omega $ and T are the geometrical factor and the live time respectively, $\Delta E$ denotes the energy bin width, $E$ is the geometric mean of the lower and upper bounds of the bin~\cite{Maurino}, 
$N(E)$ the bin content in the unfolded distribution,
$\varepsilon (E)$ the total selection efficiency (Fig.~S3 of the SM~\cite{PRL-SM}), 
$U()$ the unfolding procedure operator,
$N_{obs}(E_{\rm TASC})$ the bin content of observed energy distribution (including background),
and $N_{bg}(E_{\rm TASC})$ the bin content of background events in the observed energy distribution.
In the energy range between $ 10^2 $ and $ 10^3 $~GeV of $ E_{\mathrm{TASC}}$ the background fraction is $N_{bg}/N_{obs} \sim1\%$. Starting from $ 10^3 $~GeV it increases up to 10\% at $ 10^4 $~GeV.

\section{Systematic Uncertainties}
The most important sources of systematics uncertainties in the nickel analysis are due to the MC model and event selection at high energy.
The systematic error related to charge identification was
studied by varying the semi-minor and major axes of the elliptical selection
up to $ \pm $15\% corresponding to a variation of charge selection efficiency of $ \pm 17\%$.
The result was an (energy bin dependent) flux variation lower than 4\% below 100~GeV$ /n $ and increasing to~$ \sim$8\% at 200~GeV$ /n $.
A comparison between different MC algorithms  is
in order as it is not possible to validate the MC simulations with beam test data at high energy. A comparative study of key distributions was carried out with EPICS and GEANT4 showing that the respective total
selection efficiencies for Ni are in agreement within $ \sim$3\%
over the whole energy range (Fig.~S3 of the SM~\cite{PRL-SM}). 
The difference between the two energy response matrices is contained between -5\% and +5\%. 
The resulting fluxes show a difference around $ \sim $5\% below 40~GeV$ /n $ and less than $ \sim$10\%  in the 100--200~GeV$ /n $ region.

The uncertainty on the energy scale correction is $\pm2$\% and depends on the accuracy of the beam test calibration. It causes a rigid shift  of the flux ($ \pm 4\% $) above 30~GeV$ /n $, not affecting the spectral shape.
As  the  beam  test  model  was  not identical  to  the  instrument  in flight~\cite{CALET-PROTON},  the  difference in the spectrum ($ \pm 5\% $ up to 140~GeV$ /n $) obtained with either configuration was modeled and included in the systematic error.

The uncertainties due to the unfolding procedure  were evaluated with different response matrices computed by varying the spectral index (between -2.9 and -2.2) of the MC generation spectrum.

As the trigger threshold is much smaller than the energy of a non interacting nickel nucleus, the HE trigger efficiency is close to 100\% in the whole energy range with a negligible contribution to the systematic error. The fraction of interactions (Fig.~S5 of the SM~\cite{PRL-SM}) in the CHD, and above it, was checked by comparing the MC and the FD as explained in the SM. The contribution due to a shower event cut, rejecting non interacting particles (4\% around 10 GeV and $  2 $\% above), was evaluated and included in the systematic uncertainties.   

Possible inaccuracy of track reconstruction could affect the determination of the geometrical acceptance. The contamination due to off-acceptance events that are erroneously reconstructed inside the fiducial acceptance
was estimated by MC to be $\sim$1\% at 10~GeV$/n$ while decreasing to less than $0.1\%$ above 60~GeV$/n$. 
The systematic uncertainty on the tracking efficiency is negligible~\cite{CALET-CO}. A different tracking procedure, described in Ref.~\cite{akaike2019}, was also used to study possible systematic uncertainties in tracking efficiency. The result is consistent with the Kalman filter algorithm. 
The systematic error related to background contamination is assessed by varying  the contamination level by as much as $ \pm 50\%$. The result was a flux variation around 1\% below 100~GeV$ /n $, increasing to 3\% at 200~GeV$ /n $.

The systematic error related to the atomic mass of nickel isotopes composition reduces the normalization by $ 2.2\% $. 
Additional energy-independent systematic uncertainties affecting the flux normalization include live time (3.4\%), long-term stability ($<2.7\%$) and geometrical factor ($ \sim$1.6\%), as detailed in the SM of Ref.~\cite{CALET-ELE2017}. 
The energy dependence of all systematic errors for nickel analysis is shown in  Fig.~S8 of the SM~\cite{PRL-SM}. 
The total systematic error is computed as the sum in quadrature of all the sources of systematics in each energy bin. 

\section{Results}
\begin{figure} [!htb] \centering
	\hspace*{-5mm}
	\includegraphics[width=1.05\hsize]{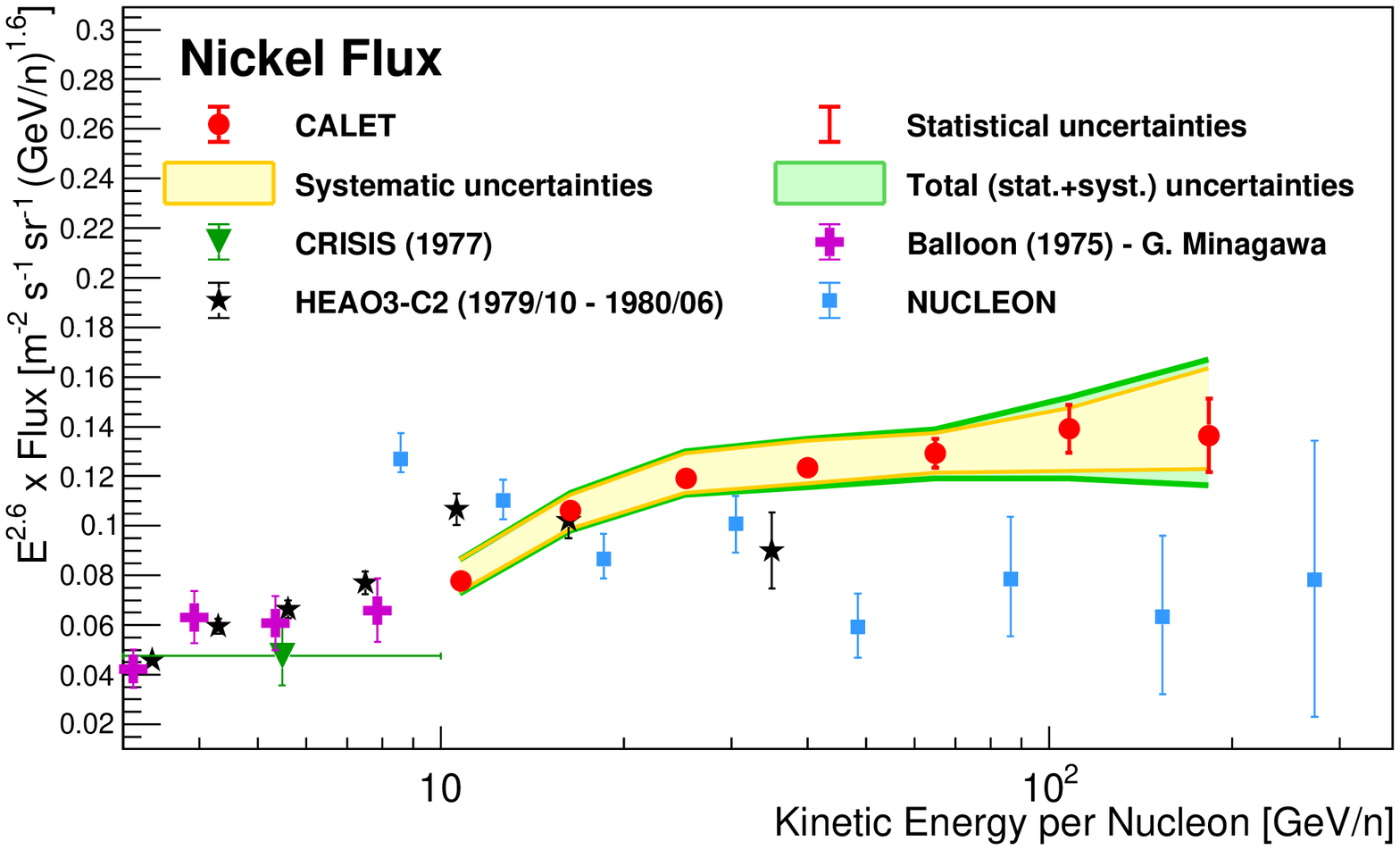}
	\caption{\scriptsize CALET nickel flux (multiplied by $E^{2.6}$) as a function of kinetic energy per nucleon. Error bars of the CALET data (red) represent the statistical uncertainty only, the yellow band indicates the quadrature sum of systematic errors, while the green band indicates the quadrature sum of statistical and systematic errors. Also plotted are the measurements from  Balloon 1975~\cite{Minagawa81}, CRISIS~\cite{Young81}, HEAO3-C2~\cite{HEAO3-Ni} and NUCLEON~\cite{NUCLEON-Ni}. This figure is reproduced and enlarged in Fig.~S9 of the SM~\cite{PRL-SM}.} 
	\label{fig:flux}
\end{figure}

The nickel differential spectrum in kinetic energy per nucleon measured by CALET in the energy range from 8.8 to 240~GeV$/n$ is shown in Fig.~\ref{fig:flux}, where current uncertainties including statistical and systematic errors are bounded within a green band. 
The CALET spectrum is compared with the results  from Balloon 1975~\cite{Minagawa81}, CRISIS~\cite{Young81}, HEAO3-C2~\cite{HEAO3-Ni} and NUCLEON~\cite{NUCLEON-Ni}. 
The nickel flux measurements with CALET are tabulated in Table~I of the SM~\cite{PRL-SM} where statistical and systematic errors are also shown. 
CALET and HEAO3-C2 nickel spectra have similar flux normalization in the common interval of energies. CALET and NUCLEON differ in the shape although the two measurements show a similar flux normalization at low energy.  

\begin{figure} \centering
	\hspace*{-5mm}
	\includegraphics[width=1.05\hsize]{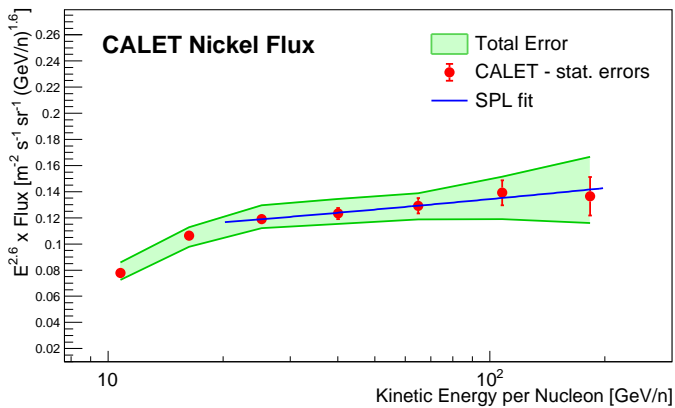}
	\caption{\scriptsize Fit of the CALET nickel energy spectrum to an SPL function (blue line) in the energy range [20, 240]~GeV$ /n $.
		The flux is multiplied  by $E^{2.6} $ where $ E $ is the kinetic energy per nucleon. The error bars are representative of purely statistical errors.
	}
	\label{fig:Fefit}
\end{figure}

Figure~\ref{fig:Fefit} shows a fit to the CALET nickel flux with a single power law function (SPL)
\begin{equation}
\Phi(E) = C\, \left(\frac{E}{\text{1 GeV}/n} \right)^{\gamma}
\label{eq:SPL}
\end{equation}
where $ \gamma $ is the spectral index and $ C $ is the normalization factor. 

\noindent

\begin{figure}[!htb]
	\centering
	\includegraphics[width=\hsize]{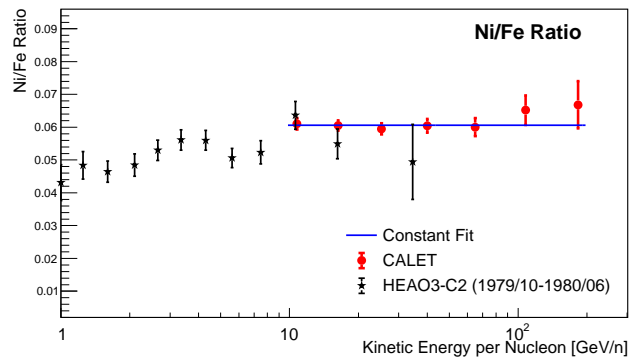}
	\caption{\scriptsize Nickel to iron flux ratio measured with CALET (red points). The errors bars are representative of statistical errors only. Data are fitted with a constant function giving Ni/Fe = 0.061 $ \pm $ 0.001. Also plotted is the result from HEAO3-C2~\cite{HEAO3-Ni}. 
		\label{fig:NiFeRatio}}
\end{figure}

The fit is performed from 20~to 240~GeV$/n$ and gives $\gamma = -2.51 \pm 0.04 (\mathrm{stat})  \pm  0.06 (\mathrm{sys})$ with a $\chi^2/$d.o.f.~=~0.3/3. Below 20 GeV$ /n$ the observed Ni flux softening is similar to the one found for iron and lighter primaries.
To better understand the nickel spectral behavior we report also the nickel to iron ratio as a function of kinetic energy per nucleon (see Fig.~\ref{fig:NiFeRatio}). Our measure extends the results of previous experiments (i.e. HEAO3-C2) up to 240~GeV$/n$. The fit, performed from 8.8~to 240~GeV$/n$, gives a constant value of $0.061 \pm 0.001$(stat) with  the $ \chi^2/$d.o.f. = 2.3/6.

The experimental limitations of the present measurement (i.e. low statistics as well as large systematic errors for the highest energy bins) do not yet allow one to test the hypothesis of a spectral shape different from a single power law in the region above 20 GeV$ /n $. As a matter of fact, current expectations (e.g.,~\cite{Thoudam,Tomassetti}) for a detectable spectral hardening of nickel are still under debate. 

\section{Conclusion}
 In this paper, based on 67 months of observations with CALET on the ISS, we report for the first time a measurement of the energy spectrum of nickel over an extended energy range up to 240~GeV$ /n $ and with a significantly better precision than most of the existing measurements. 
The nickel spectrum behavior below 20 GeV/n is similar to the one observed for iron and lighter primaries. Above 20~GeV$ /n $, our present observations are consistent with the hypothesis of an SPL spectrum up to 240~GeV$ /n $.
Beyond this limit, the uncertainties given by our present statistics and large systematics do not allow us to draw a significant conclusion on a possible deviation from a single power law. An SPL fit in this region yields a spectral index value $ \gamma = -2.51 \pm 0.07 $.
The flat behavior of the nickel to iron ratio suggests that the spectral shapes of Fe and Ni are the same within the experimental  accuracy. This  suggests a similar acceleration and propagation behavior as expected from the small difference~in atomic number and weight between  Fe and Ni nuclei.
An extended data set, as expected beyond the 67 month period of continuous observations accomplished so far, will not only improve the most important statistical limitations of the present measurement, but also our understanding of the instrument response in view of a further reduction of systematic uncertainties.

\section{Acknowledgments}
\begin{acknowledgments}
We gratefully acknowledge JAXA's contributions to the development of CALET and to the operations aboard the JEM-EF on the International Space Station.
We also wish to express our sincere gratitude to ASI (Agenzia Spaziale Italiana) and NASA for their support of the CALET project.
This work was supported in part by JSPS KAKENHI Grant No.~26220708, No.~19H05608, No.~17H02901, No.~21K03592, and No.~20K22352 and by the 
MEXT-Supported Program for the Strategic Research Foundation at Private Universities (2011-2015)
(No.~S1101021) at Waseda University.
The CALET effort in Italy is supported by ASI under agreement No.~2013-018-R.0 and its amendments.
The CALET effort in the United States is supported by NASA through Grants No.~80NSSC20K0397, No.~80NSSC20K0399, and No.~NNH18ZDA001N-APRA18-0004. 
\end{acknowledgments}

\nocite{*}

\providecommand{\noopsort}[1]{}\providecommand{\singleletter}[1]{#1}%

\clearpage
\widetext

\setcounter{equation}{0}
\setcounter{figure}{0}
\setcounter{table}{0}
\setcounter{page}{1}
\makeatletter
\renewcommand{\theequation}{S\arabic{equation}}
\renewcommand{\thefigure}{S\arabic{figure}}
\renewcommand{\bibnumfmt}[1]{[S#1]}
\renewcommand{\citenumfont}[1]{S#1}
\begin{center}
	\textbf{\large Direct Measurement of the Nickel Spectrum in Cosmic Rays in the Energy Range \\ from 8.8 GeV/${n}$ to 240 GeV/${n}$
		with CALET on the International Space Station \\
		\vspace*{0.5cm}
		SUPPLEMENTAL MATERIAL}	\\
	\vspace*{0.2cm}
	(CALET collaboration) 
\end{center}
\vspace*{1cm}
Supplemental material relative to ``Direct Measurement of the Nickel Spectrum in Cosmic Rays in the Energy Range from 8.8 GeV/${n}$ to 240~GeV/${n
}$ with CALET on the International Space Station with the Calorimetric Electron Telescope''.
\vspace*{1cm}

\clearpage
\section{Additional information on the analysis}

\begin{figure}[h!]  \centering
	\subfigure[]{\includegraphics[width=0.8\hsize]{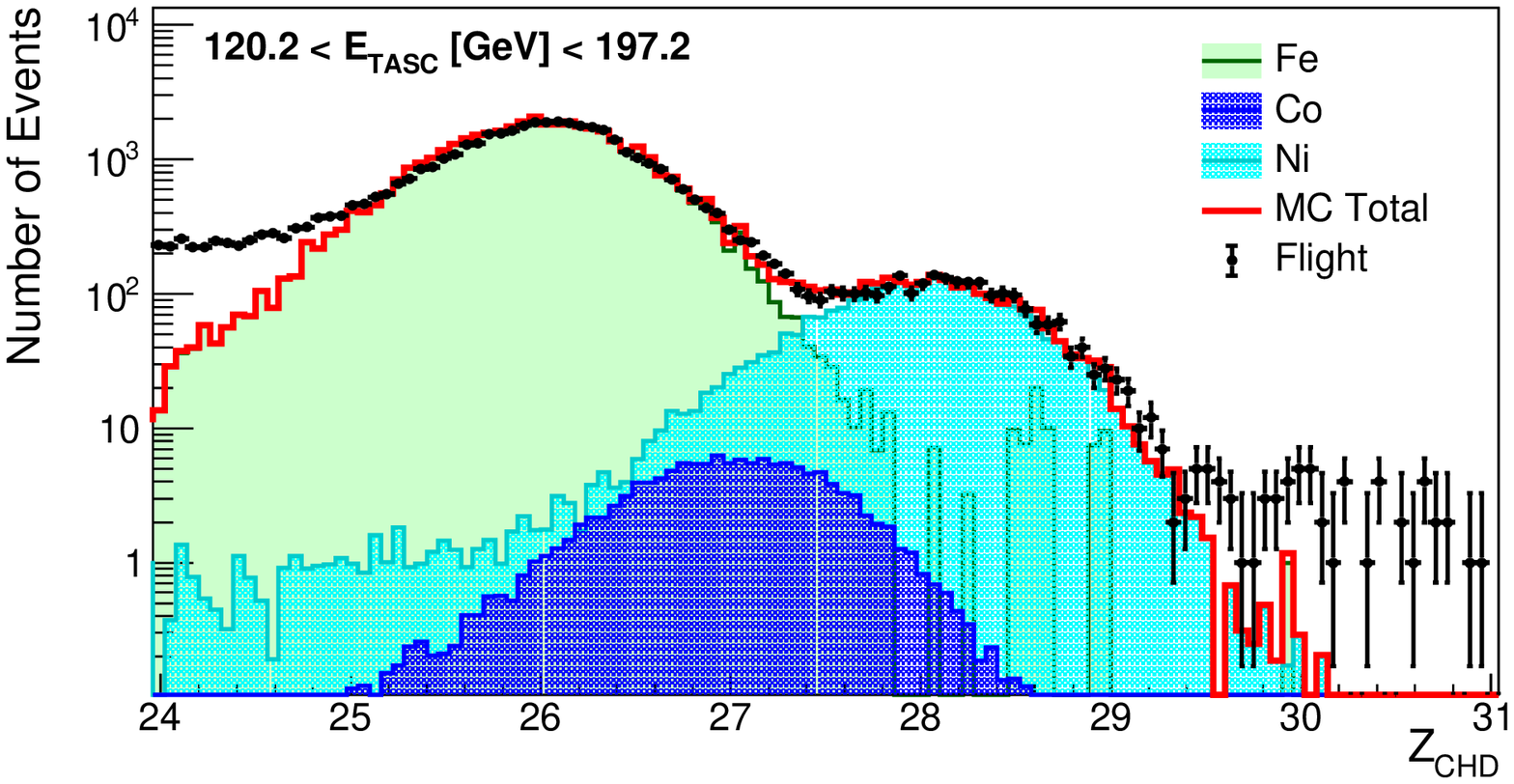}}
	\subfigure[]{\includegraphics[width=0.8\hsize]{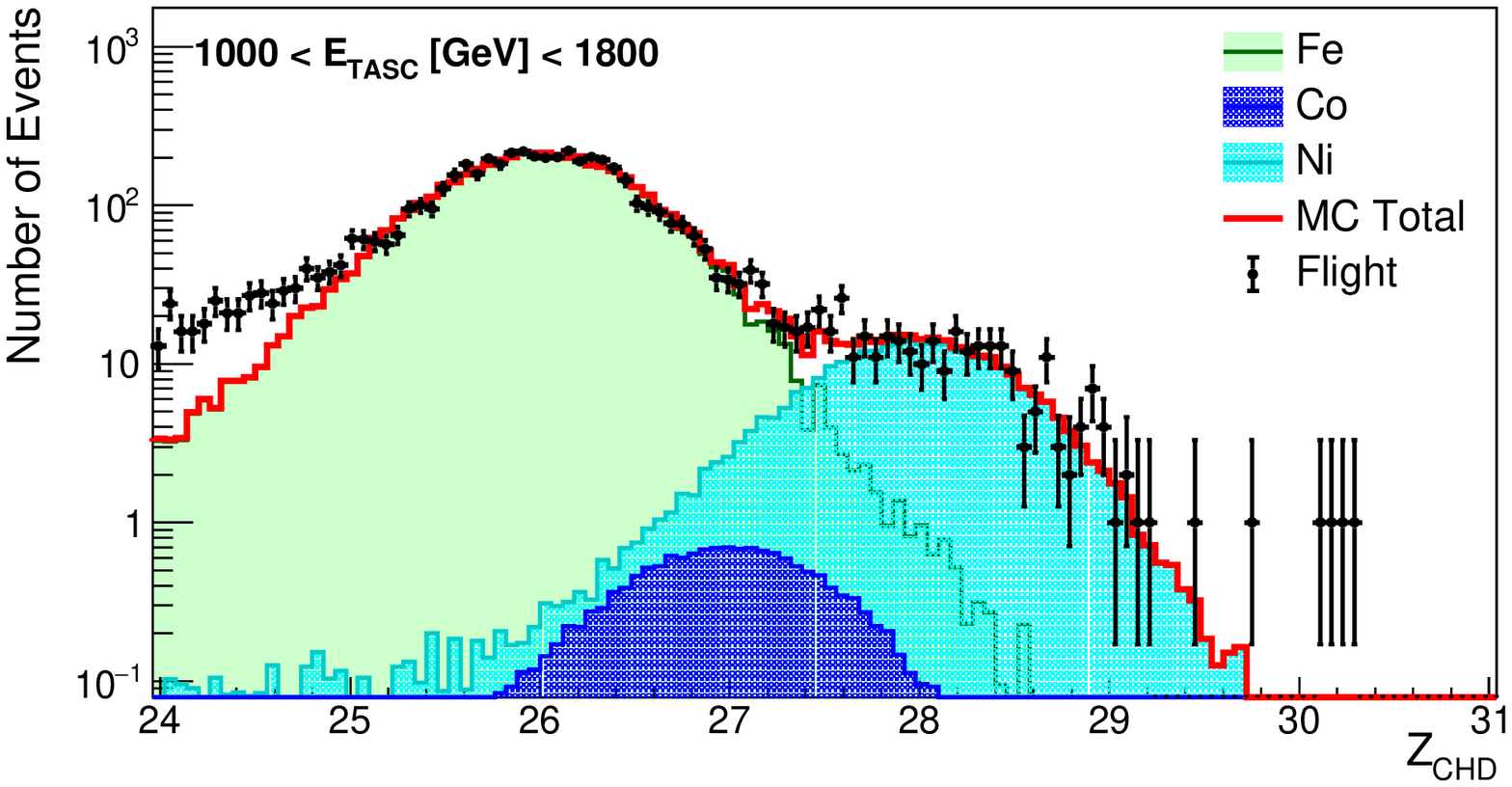}}
	\caption{Charge distributions from the combined charge measurement of the two CHD layers in the elemental region between Cr and Ga. 
		Events are selected with $120.2 < E_{\rm TASC} < 197.2$  GeV in (a) and $1000 < E_{\rm TASC} < 1800$ GeV in (b). Flight data, represented by black dots, are compared with Monte Carlo samples including iron, cobalt and nickel. Chromium, manganese and nuclei with $ Z\geq 29 $ are not included in MC because their contamination to nickel data is negligible. }
	\label{fig:Z_CHD_Niregion_SM}
\end{figure}

$\bf{Charge \, measurement. \,}$ 
The particle's charge $ Z $ is reconstructed from the ionization deposits in the CHD paddles traversed by the incident particle. Either CHD layer provides an independent $ dE/dx $ measurement which is corrected for the track path length and for the quenching effect in the scintillator's light yield  as a function of $ Z^2 $~(Refs.~\cite{GSI-SM, CALET-CO-SM}).  \\
In Fig.~\ref{fig:Z_CHD_Niregion_SM} inclusive distributions of measured charges from flight data (FD) are compared, in two different energy bins, with Monte Carlo (MC) simulations from EPICS.

\noindent

\begin{figure}[!htb]
	\begin{center}
		\includegraphics[scale=0.7]{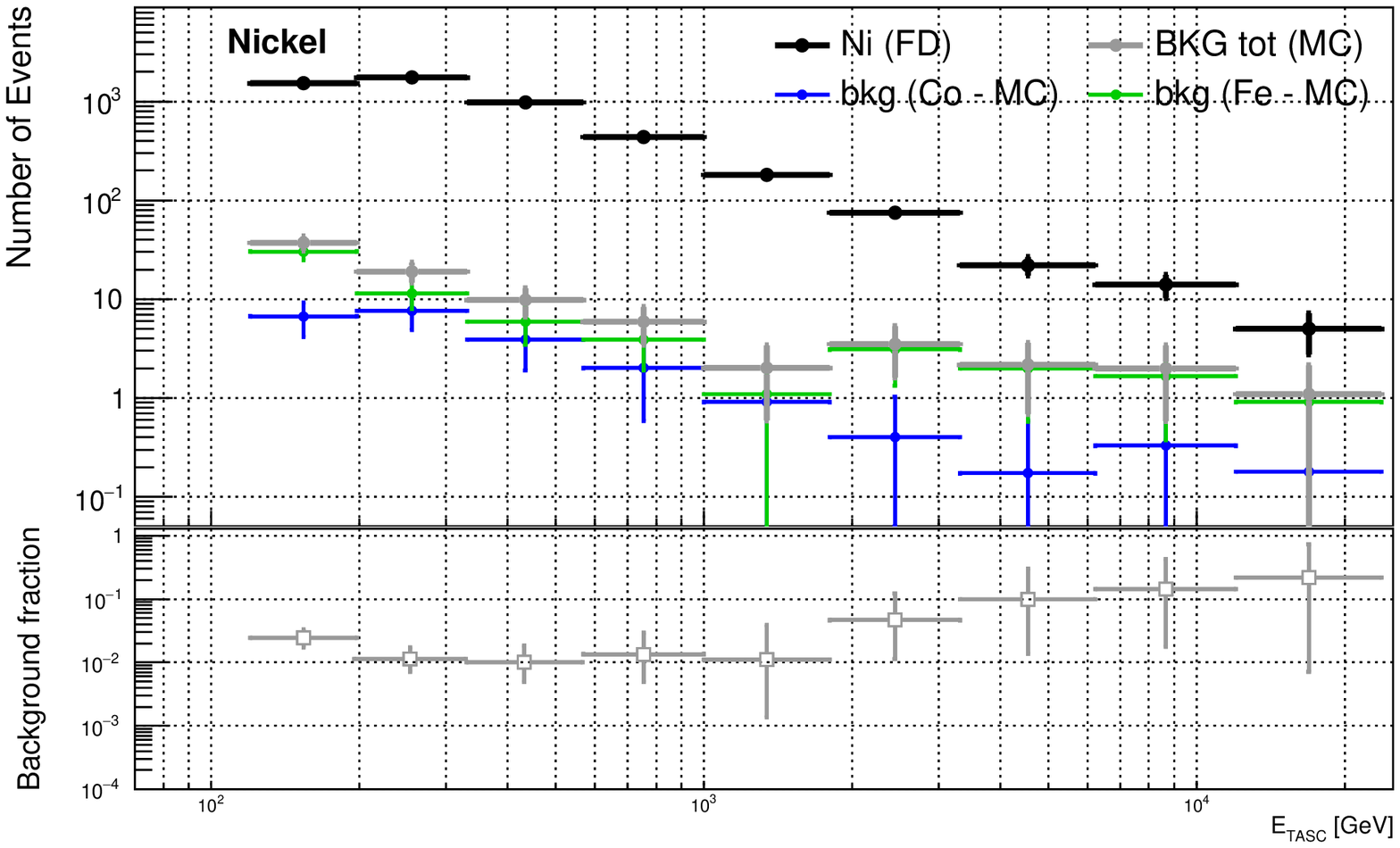}
		\caption{Top panel: Differential distributions of the number of events in a given bin of calorimetric energy ($E_{\rm TASC}$ in GeV) for selected nickel events in flight data (black dots) before the unfolding procedure and with background events from iron and cobalt nuclei. Bottom panel: Contamination from Fe and Co obtained with the MC. The Monte Carlo events are weighted with a factor to reproduce a single power law spectrum with spectral index -2.6, and event selection is the same as for flight data. The resulting elemental charge distribution in each observed energy bin has been normalized to match the CHD charge distribution of flight data. The number of contaminant events is calculated by integration of all the MC events retained by the nickel charge selection.
		}
		\label{fig:TASCedepC}
	\end{center}
\end{figure}

\noindent
$\bf{Background \, contamination.}$ 
Background contamination of nickel from neighbor elements is relatively small between 100~GeV and 1~TeV mainly due to iron and secondly to cobalt. Above 1~TeV the iron contribution becomes important and exceeds 10\% above 10~TeV. 
The distribution of nickel candidates and its contaminants is shown in Fig.~\ref{fig:TASCedepC} as a function of the TASC energy. The total contamination  (estimated by fitting  a constant  up to 1.7~TeV and a first order logarithmic   polynomial up to 20~TeV) was subtracted from the flux as explained in the main body of the paper.
To evaluate the systematic error related to the background, the constant parameter of the fit was varied by as much as twice its error. This resulted  into a   $ \pm 50\% $ variation of the contamination with a negligible change of the flux (see main body). 
\newline
\indent

\begin{figure}[!htb] \centering
	\includegraphics[width=0.75\hsize]{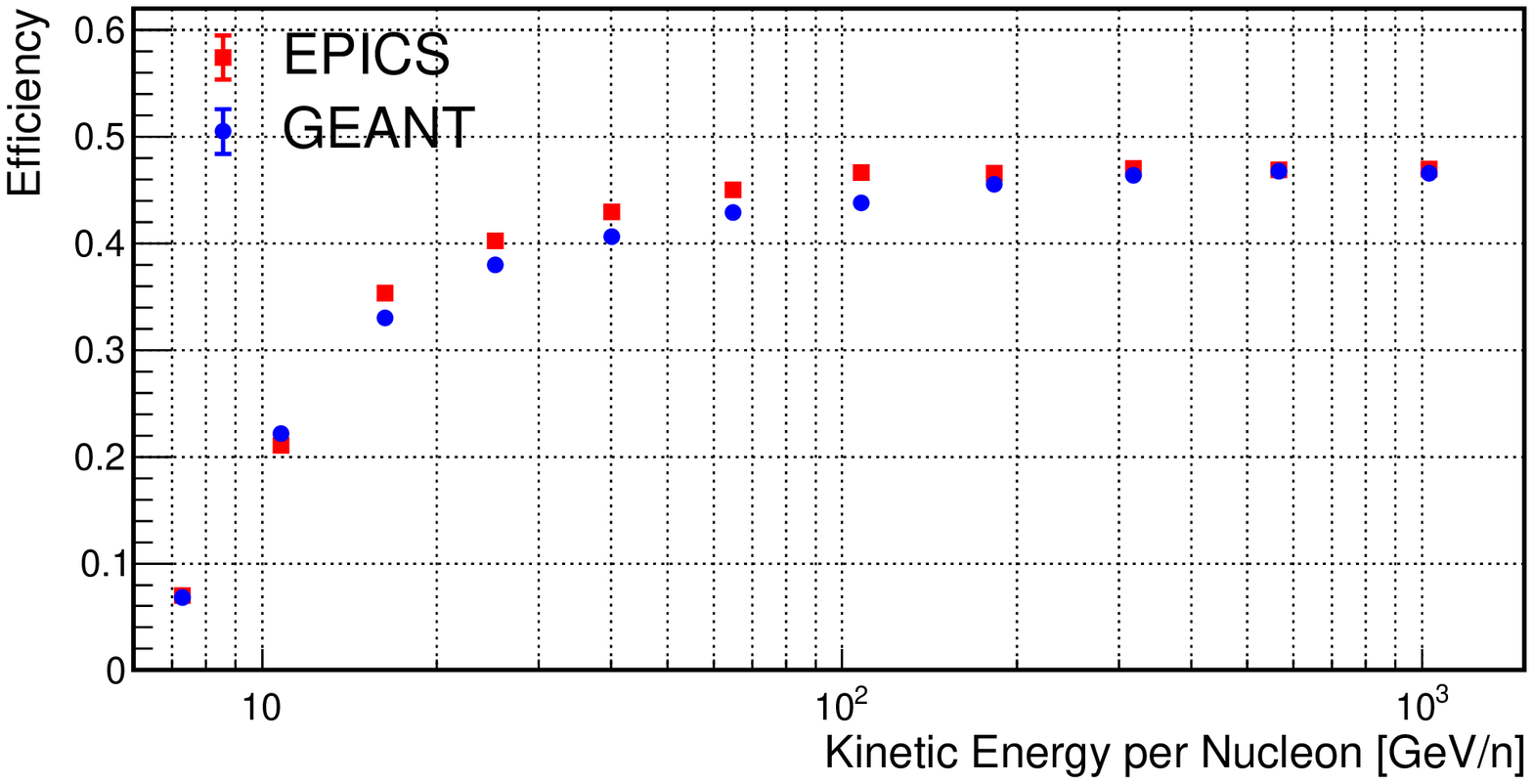}
	\caption{Total selection efficiency for nickel events as estimated with EPICS (red filled circles) and GEANT4 (blue filled circles) simulations.
	}
	\label{fig:tot_eff}
\end{figure}

\noindent
$\bf{Efficiencies}.$ The total efficiency and relative efficiencies (i.e., the efficiency of a given cut normalized to the previous cut) were studied extensively over the whole energy range covered by the nickel flux measurement. 
The total selection efficiency from EPICS (red filled circles) and GEANT4 (blue filled circles) are shown in Fig.~\ref{fig:tot_eff} as a function of total particle kinetic energy per nucleon. The above efficiencies were validated by comparing distributions relevant to the selection of events, and obtained from flight data, with the same distributions generated by EPICS or GEANT4. \\

\begin{figure}[!htb] \centering
	\includegraphics[width=\hsize]{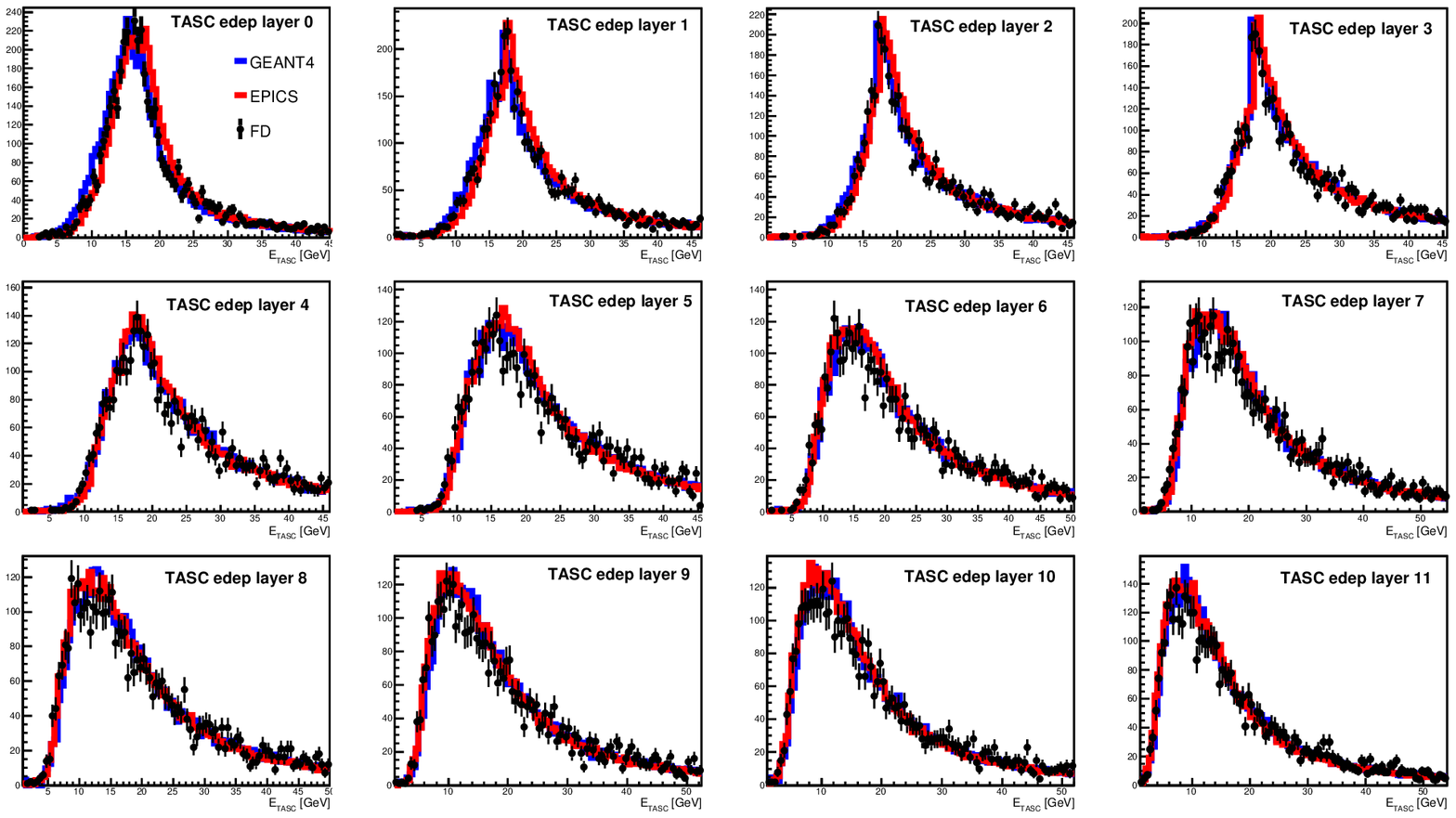}
	\caption{
		Energy observed in each of the 12 layers of the TASC (black points) for the final sample of nickel candidates from flight data. It is compared with pure samples of nickel simulated by EPICS (red) or GEANT4 (blue). 
	}
	\label{fig:TASC_layers}
\end{figure}

\noindent
$\bf{TASC \, energy \, deposits. \,}$  An example is given in Fig.~\ref{fig:TASC_layers} where the total energy observed in each layer of the TASC (black points) was plotted using the final sample of nickel candidates, marginally contaminated by a residual background due to elements with atomic number close to nickel. Compared with pure nickel samples simulated by EPICS (red) or GEANT4 (blue), the distributions from the two MC were found to be consistent with each other and in fair agreement with flight data.\\

\begin{figure}[!htb] \centering
	\includegraphics[scale=0.7]{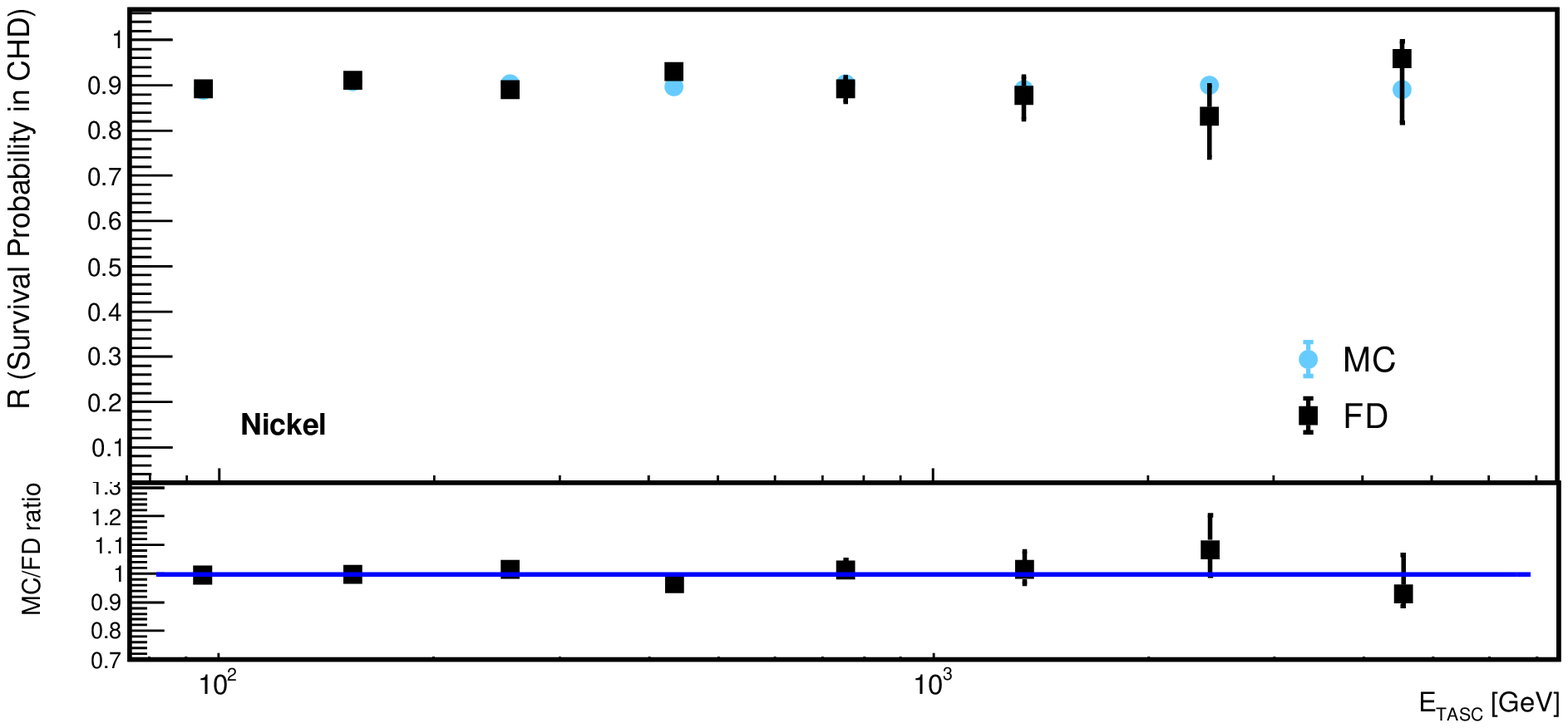}
	\caption{Top panel: nickel survival probability R as a function of energy for events crossing both layers of the CHD with flight data (black filled squares) and EPICS (blue filled circles); Bottom panel: ratio of $R_{MC} / R_{FD}$ with the MC and flight data, respectively, fitted with a constant and consistent with unity within the error.
	}
	\label{fig:survival_prob}
\end{figure}

\noindent
$\bf{Interactions \, in\, the \, instrument}.$ The amount of instrument material above the CHD is very small and well known. The largest significant contribution is limited to a 2 mm thick Al cover placed on top of the CHD. This amounts to $\sim2.2\%$ radiation length and $5\times10^{-3}$ proton interaction lengths ($\lambda_{I} $). The material description in the MC is very accurate and derived directly from the CAD model.  As CALET is sitting on the JEM external platform of the ISS, no extra material external to CALET is normally present within the acceptance for the flux measurement. 
However, occasional obstructions caused by the ISS robotic arm operations may temporarily affect the field of view (FOV). It was checked that removing those rare periods from the flux results in a negligible difference in shape and normalization~($ < $1\%). 
\newline
\indent
MC simulations were used to evaluate the nickel survival probability after traversing both layers of the CHD and the material above. In order to check its consistency with flight data, the ratio R = (CHDX $\&$ CHDY) / CHDX (i.e. the fraction of nickel candidates tagged by both CHD layers among those detected by the top charge detector alone) was plotted, as a function of the TASC energy, for selected nickel candidates with measured charge in the range 27.7~-~28.3 charge units. 
In the upper panel of Fig.~\ref{fig:survival_prob}, R is shown in 8 bins of the TASC energy for both MC ($R_{MC}$) and FD ($R_{FD}$) with an average value around $90\%$ and a flat energy dependence. 
The $R_{MC} / R_{FD} \, \it{double \, ratio}$  (lower plot) shows a good level of consistency between the MC and flight data, within the errors.
The total loss ($\sim 10\%$) of nickel events interacting in the upper part of the instrument was taken into account in the total efficiency and its uncertainty included in the systematic error.
\newline

\begin{figure}[!htb]
	\centering
	\includegraphics[width=\hsize]{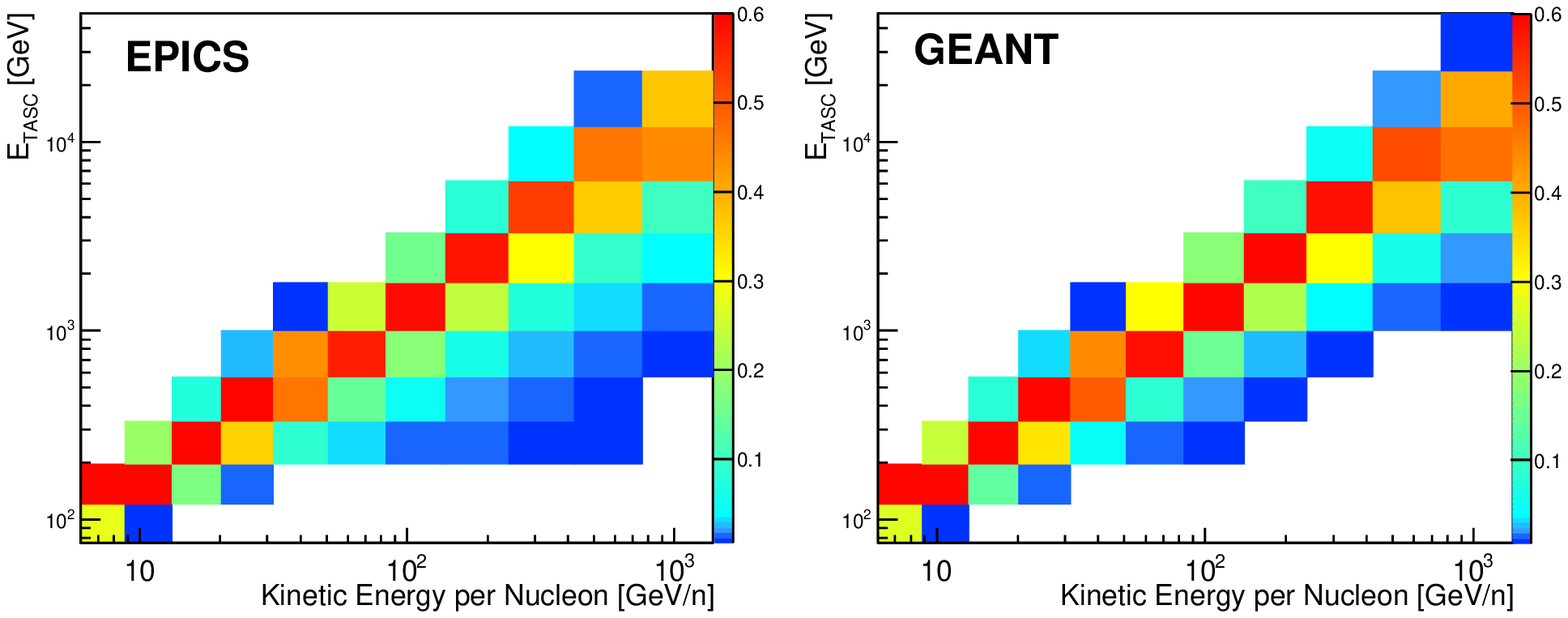}
	\caption{Response matrix for nickel derived from the MC simulations of the CALET flight model by applying the same selection as for flight data. The array is normalized so that the color scale is associated with the probability that nickel candidates, in a given bin of particle kinetic energy, migrate to different intervals of $ E_\mathrm{TASC} $\label{fig:UFmatrix}. Left: EPICS; Right: GEANT4.
	} 
\end{figure}

\noindent
$\bf{Calorimetric \, energy, \, bin \, size\,,and \, unfolding.}$
The energy response of the TASC was studied with MC simulations and compared with the results of measurements of the total particle energy vs beam momentum carried out at CERN.  During one of the beam test campaigns of CALET at the SPS with an extracted primary beam of $^{40}$Ar (150~GeV$ /c/n $), beam fragments were generated from an internal target and guided toward the instrument along a magnetic beam spectrometer that provided an accurate selection of their rigidity and A/Z ratio. The relation between the observed TASC energy and the primary energy was measured for several nuclei up to the highest available energy (6~TeV total particle energy in the case of $^{40}$Ar). After an offline rejection of a very small amount of beam contaminants from the data, the shape of the TASC total energy was found to be consistent with a Gaussian distribution (Ref.~\cite{akaike2015-SM}). 
\newline
\indent
The correlation matrix used for the unfolding was derived from the simulations, using two different MC codes EPICS and GEANT4, and applying the same selection cuts as in the FD analysis. The normalized unfolding matrices obtained from EPICS and GEANT4 are shown in Fig.~\ref{fig:UFmatrix} where the color scale indicates the probability for a nickel candidate, with a given primary  energy, of depositing energy in different intervals of $ \mathrm{E_{TASC} }$.  

\begin{figure}[!htb] \centering
	\includegraphics[width=0.8\hsize]{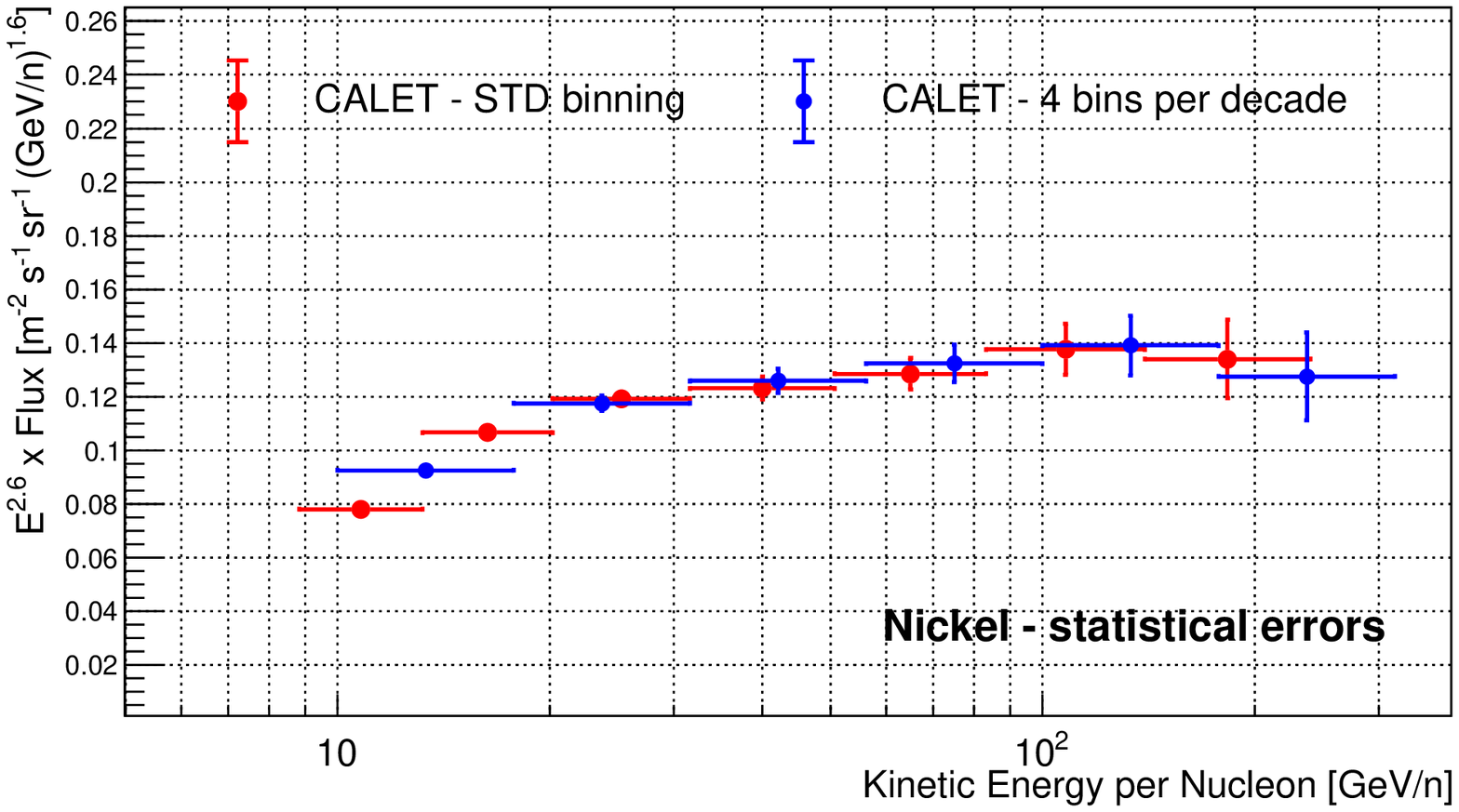}
	\caption{CALET nickel flux with standard binning (red circles) and 4 bins/decade (blue circles). The vertical error bars are representative of purely statistical errors whereas the horizontal ones indicate the bin width.
	}
	\label{fig:flux_with_3_binnings}
\end{figure}

A standard binning scheme, consisting of smoothly enlarged bins, was chosen in order to balance the statistical population at high energy without using too large 
consecutive bins. 
In order to investigate the effect of the binning choice, different binning configurations (3 bins per decade, 4 bins per decade and 5 bins per decade) were tested, obtaining similar smearing matrices and almost identical behavior in the final flux. 
As an example, the difference between the flux with the standard binning and with 4 bins per decade is shown in Fig.~\ref{fig:flux_with_3_binnings}. 
\newline
\indent

\noindent

\begin{figure}[hbt!] \centering
	\includegraphics[scale=0.75]{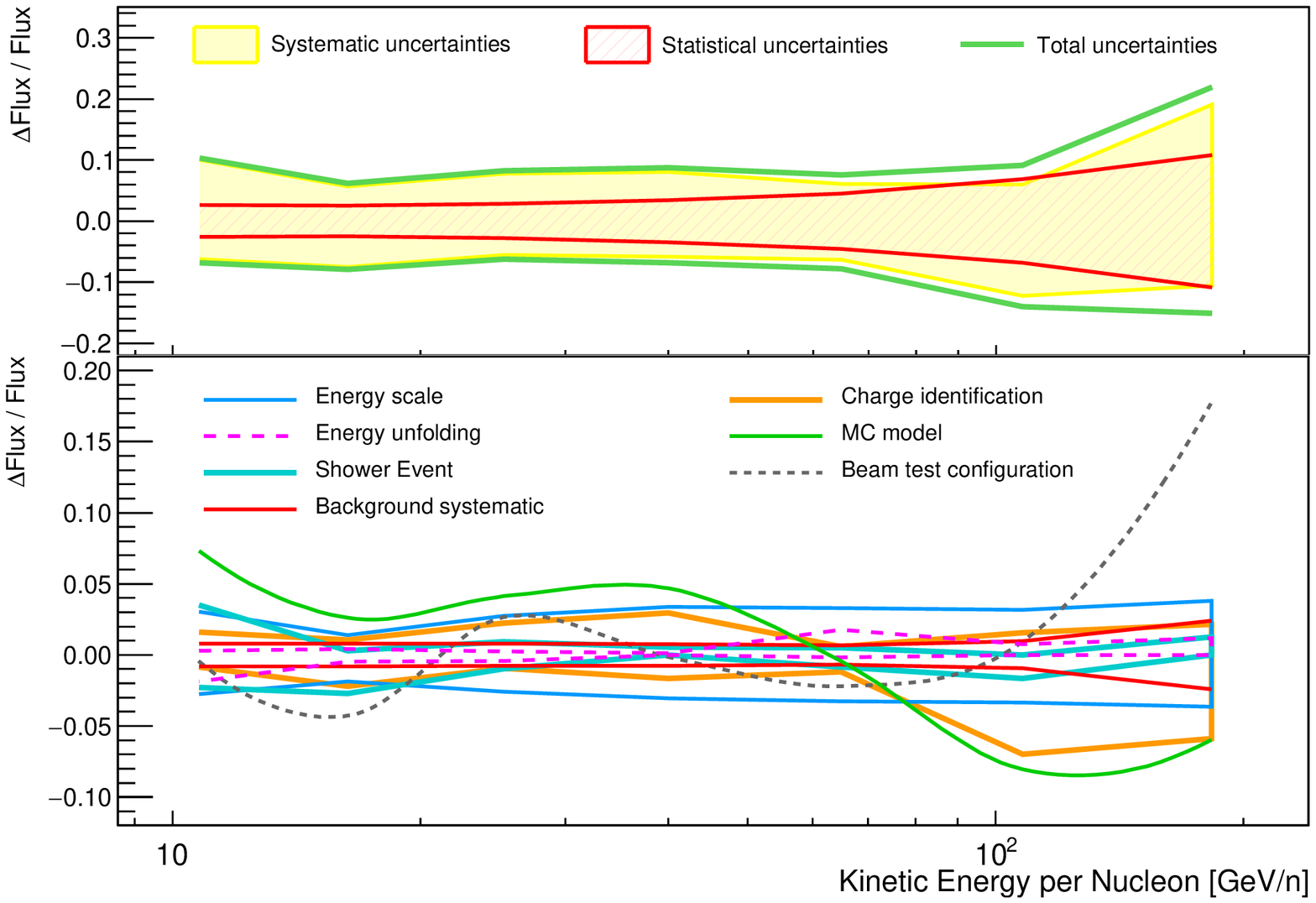}
	\caption{Energy dependence (in GeV/nucleon) of systematic uncertainties (relative errors) for nickel.  In the upper panel, the band bounded by the red lines represents the statistical error. The yellow band shows the sum in quadrature of all the sources of systematics including energy independent ones. The green lines represent the sum in quadrature of statistical and total systematic uncertainties. In the bottom panel, a detailed breakdown of systematic energy dependent errors, stemming from charge identification, MC model, energy scale correction, energy unfolding, beam test configuration and shower event shape is shown. 
	}
	\label{fig:sys_all}
\end{figure}

\noindent
$\bf{Energy \, dependent \, systematic \, errors.}$
A breakdown of energy dependent systematic errors stemming from several sources (as explained in the main body of the paper) and including selection cuts, charge identification, MC model, energy scale correction, energy unfolding, beam test configuration and shower event shape is shown in Fig.~\ref{fig:sys_all} as a function of kinetic energy per nucleon. 
\newline
\indent

\begin{figure}[!htb]
	\centering
	\includegraphics[width=0.8\hsize]{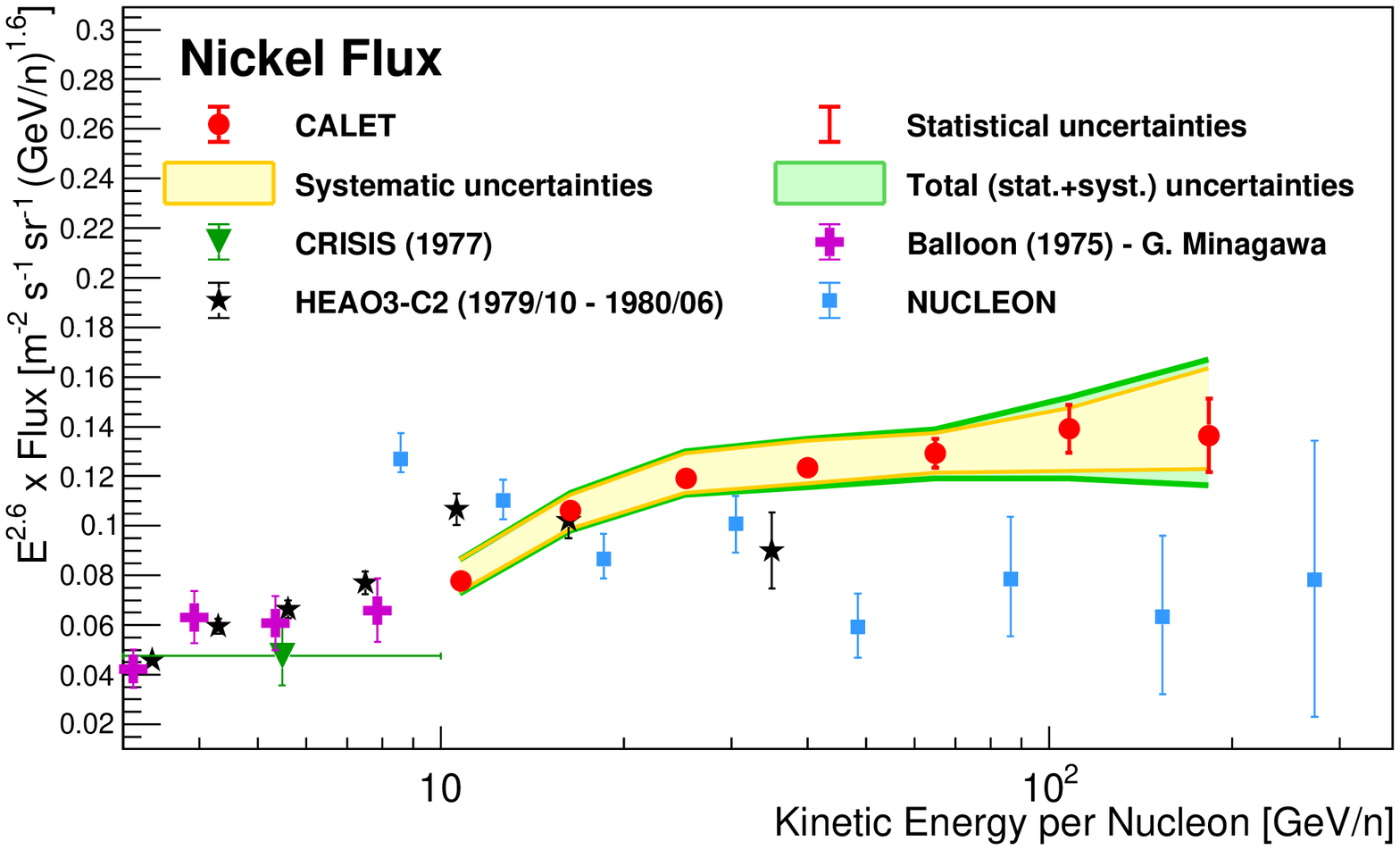}
	\caption{CALET nickel flux as a function of kinetic energy per nucleon in GeV (with multiplicative factor $E^{2.6}$). The error bars of the CALET data (red filled circles) represent the statistical uncertainty only. The yellow band indicates the quadrature sum of systematic errors, while the green band indicates the quadrature sum of statistical and systematic errors. Also plotted are the data points from Balloon 1975~\cite{Minagawa81-SM}, CRISIS~\cite{Young81-SM},  HEAO3-C2~\cite{HEAO-SM} and NUCLEON~\cite{NUCLEON-Ni-SM}. \label{fig:FluxSM}}
\end{figure}

\noindent
$\bf{Nickel \, flux \, normalization \,and \, spectral \, shape.}$
The CALET nickel flux and a compilation of available data are shown in Fig.~\ref{fig:FluxSM}, as an enlarged version of Fig.~2 in the main body of the paper.
\newline
\indent\

\clearpage
\renewcommand{\arraystretch}{1.25}
\begin{table*}
	\caption{Table of the CALET differential spectrum in kinetic energy per nucleon of cosmic-ray nickel. 
		The first, second, and third error in the flux are representative of the statistical uncertainties, systematic uncertainties in normalization, and energy dependent systematic uncertainties, respectively.
		\label{tab:Cflux}}
	\begin{ruledtabular}
		\begin{tabular}{c c c c}
			Energy Bin [GeV$ /n $] & Flux [m$^{-2}$sr$^{-1}$s$^{-1}$(GeV$/n)^{-1}$]   \\
			\hline
			8.8 -- 13.2 & $( 1.61 \, \pm 0.04 \,  _{- 0.08 }^{+ 0.07} \, _{- 0.07 }^{+ 0.14 }) \times 10^{- 4 }$ \\
			13.2 -- 20.2 & $( 7.48 \, \pm 0.19 \,  _{- 0.38 }^{+ 0.35} \, _{- 0.45 }^{+ 0.24 }) \times 10^{- 5 }$ \\
			20.2 -- 31.6 & $( 2.69 \, \pm 0.08 \,  _{- 0.14 }^{+ 0.12} \, _{- 0.06 }^{+ 0.19 }) \times 10^{- 5 }$ \\
			31.6 -- 50.7 & $( 8.41 \, \pm 0.29 \,  _{- 0.43 }^{+ 0.39} \, _{- 0.25 }^{+ 0.59 }) \times 10^{- 6 }$ \\
			50.7 -- 83.2 & $( 2.51 \, \pm 0.11 \,  _{- 0.13 }^{+ 0.12} \, _{- 0.12 }^{+ 0.09 }) \times 10^{- 6 }$ \\
			83.2 -- 139,6 & $( 7.23 \, \pm 0.50 \,  _{- 0.37 }^{+ 0.33} \, _{- 0.85 }^{+ 0.20 }) \times 10^{- 7 }$ \\
			139.6 -- 239.9 & $( 1.79 \, \pm 0.19 \,  _{- 0.09 }^{+ 0.08} \, _{- 0.16 }^{+ 0.33 }) \times 10^{- 7 }$ \\
		\end{tabular}
	\end{ruledtabular}
\end{table*}
\renewcommand{\arraystretch}{1.0}

\providecommand{\noopsort}[1]{}\providecommand{\singleletter}[1]{#1}%
%

\end{document}